\documentclass{article}
\usepackage{spconf,amsmath,graphicx}
\usepackage{subfig}
\usepackage{multirow}  
\usepackage{amssymb}
\usepackage{amsmath}
\usepackage{booktabs}
\usepackage{url}

\title{TS-RIR: Translated synthetic room impulse responses for speech augmentation}
\name{Anton Ratnarajah, Zhenyu Tang, Dinesh Manocha} 
\address{University of Maryland, College Park, MD 20742, United States}
\begin{document}
%
\maketitle
\begin{abstract}
We present a method for improving the quality of synthetic room impulse responses for far-field speech recognition. We bridge the gap between the fidelity of synthetic room impulse responses (RIRs) and the real room impulse responses using our novel, TS-RIRGAN architecture. Given a synthetic RIR in the form of raw audio, we use TS-RIRGAN to translate it into a real RIR. We also perform real-world sub-band room equalization on the translated synthetic RIR. Our overall approach improves the quality of synthetic RIRs by compensating low-frequency wave effects, similar to those in real RIRs. We evaluate the performance of improved synthetic RIRs on a far-field speech dataset augmented by convolving the LibriSpeech clean speech dataset \cite{LibriSpeech} with RIRs and adding background noise. We show that far-field speech augmented using our improved synthetic RIRs reduces the word error rate by up to 19.9\% in Kaldi far-field automatic speech recognition benchmark \cite{zhenyu_low}.
\end{abstract}
\section{Introduction}
\label{sec:intro}

Far-field speech recognition is still a challenging problem because only a limited number of far-field speech corpora are available~\cite{microsoft_noise,VOiCES,acoustic_matching}. Unlike near-field speech, which is recorded close to the speaker, far-field speech may contain strong reverberation effects. The reverberation effects are associated due to various factors, including the room layout, speaker and listener positions, obstacles, and room materials. The reverberation effects can be mathematically modeled as a transfer function known as Room Impulse Response (RIR). We can augment far-field speech by convolving clean speech with an RIR and adding environmental noise with different signal-to-noise ratios \cite{ButReverb,study1}. 

The RIR can be captured from an acoustic environment using different techniques \cite{Time-Stretched-Pulses, Sine_Sweep, microsoft,rir_capture}. Recording real RIRs require a lot of human labor and special hardware. As a result, many far-field automatic speech recognition systems use synthetic RIRs for training \cite{google_speech,speech1,speech2,zhenyu_GAS}. Synthetic RIRs can be generated using physically-based acoustic simulators for different scenes \cite{study1,wave1,image_method}. The current acoustic simulators have resulted in considerable improvement in far-field speech recognition \cite{zhenyu_GAS}. However, there is still a gap between the performance of RIRs generated using acoustic simulators and the performance of real RIRs. Most commonly used acoustic simulators are unable to model all the acoustic effects in the environment, which can be captured by real RIRs. For example, ray-tracing-based acoustic simulators \cite{zhenyu_GAS} can only simulate high-frequency acoustic effects but are not accurate in terms of low-frequency effects like diffraction or interference. 


In the computer vision literature, neural networks are used to translate simple sketches to photo-realistic images \cite{domain1,domain2}. Free-hand sketches are spatially imprecise and geometrically distorted \cite{domain1}. In particular, a neural network (CycleGAN \cite{cycle-GAN}) has been proposed to translate imprecise sketches to realistic images. The goal of translation in a CycleGAN is to learn a mapping between the source domain $X$ and the target domain $Y$ in the absence of paired examples. Motivated by the performance of CycleGAN in improving imprecise images, our goal is to develop a similar approach to improve the fidelity of synthetic RIRs for automatic speech recognition (ASR) applications.  

 

{\bf Main Results:} We present a novel approach to improve the accuracy of synthetic RIRs. We design a TS-RIRGAN architecture to translate the synthetic RIR to a real RIR. TS-RIRGAN takes synthetic RIRs as 1x16384 audio samples to translate them into real RIRs and use multiple loss functions. We also perform real-world sub-band room equalization to the translated RIRs to further improve their quality. We also demonstrate the benefits of our post-processed RIRs in far-field ASR systems. Our main contributions include:-  

\begin{itemize}
    \item We present our TS-RIRGAN architecture, which is used to translate an imprecise synthetic RIR to a real RIR.
    \item We propose a scheme to further improve the wave effects of synthetic RIRs by performing sub-band room equalization.
    \item We show that, on a modified Kaldi LibriSpeech far-field ASR benchmark \cite{zhenyu_low}, far-field speech augmented using our improved RIRs outperforms the far-field speech augmented using unmodified RIRs by up to 19.9\%.
\end{itemize}

The rest of the paper is organized as follows. In Section 2 we describe different techniques to generate synthetic RIRs for ASR and other speech augmentation applications. We present our novel approach to improve synthetic RIRs in Section 3. Section 4 shows the benefit of improving synthetic RIRs in far-field ASR systems. We published our code for follow-up research~\footnote{\url{https://github.com/anton-jeran/TS-RIR}}.

\section{Related Work}
\label{sec:related}

\subsection{Synthetic RIRs and Acoustic Simulation}
There are several approaches for generating RIRs for different acoustic environments. Among the existing methods, computing RIRs for indoor or outdoor scenes by numerically solving the wave equation gives the most accurate results for a given scene \cite{wave1,liu2020sound}. However, wave-based approaches are computationally expensive and their complexity can grow as the fourth power of frequency. As a result, they are practical for lower frequencies (e.g., less than 1000Hz). GAN based RIR generators \cite{irgan, image2reverb} can be used to generate RIRs for different environments, though their accuracy can vary.

A simpler and less accurate alternative to the wave-based approach is use of geometric sound propagation techniques \cite{image_method,zhenyu_GAS}. In geometric acoustic simulators, the sound is assumed to propagate as a ray. Therefore, some low-frequency characteristics of sound waves cannot be modeled using these simulators. The ray assumption is valid when the wavelength of the sound is significantly smaller than the size of the obstacle in the environment. However, the low-frequency components are not modeled accurately, when the wavelength is large. The image method \cite{image_method} and path tracing methods \cite{zhenyu_GAS} are common geometric acoustic simulation methods. The image method models only specular reflections. Path tracing methods can model both specular and diffuse reflections.


\subsection{Techniques for improving synthetic RIR}
The geometric acoustic simulators are unable to model low-frequency wave effects such as diffraction \cite{wave-diffraction} and room resonance \cite{wave-resonance} because of the ray assumption. On the other hand, we observe a boost or diminishing effect in the frequency response at different frequency bands in real RIRs due to wave modes created by room resonance. Some methods tend to compensate for the missing room response in synthetic RIRs using a sub-band room equalization approach \cite{zhenyu_low}. 



\section{Our Approach}

\begin{figure}[t] 
	\centering
	\includegraphics[width=3.3in]{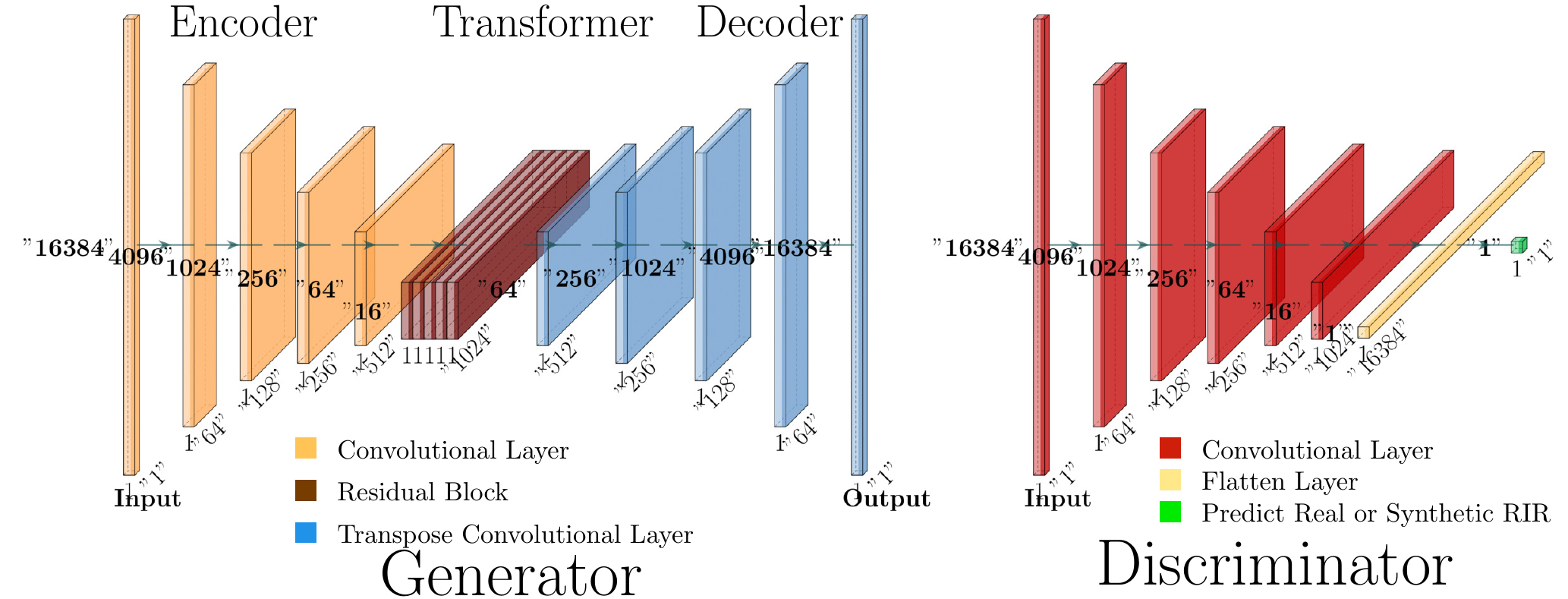}
	\caption{The architecture of the generator and discriminator of TS-RIRGAN. Our generator  takes a synthetic RIR as 1x16384 audio samples and translates it into a real RIR of the same dimension. The discriminator network discriminates between real RIRs and translated synthetic RIRs during training by maximizing the adversarial loss (Equation \ref{ad_loss}).} 
	\label{architecture}
\end{figure}

We translate synthetic RIRs to real RIRs and perform sub-band room equalization to improve the quality of synthetic RIRs. We compute the spectrogram for post-processed and real RIRs. We evaluate the quality of post-processed synthetic RIRs by comparing their mean value for a set of acoustic parameters (Table \ref{table:statistics}) and their energy distribution (Figure \ref{spectrogram}) with real RIRs. 

\subsection{Translation: Synthetic RIR $\Longrightarrow$ Real RIR}

Our TS-RIRGAN (Figure \ref{architecture}) architecture is based upon CycleGAN \cite{cycle-GAN}, and WaveGAN \cite{WaveGAN}. CycleGAN learns to translate two-dimensional images from a source domain $X$ to a target domain $Y$ using unpaired image datasets. Similarly, we design a TS-RIRGAN architecture that learns mapping functions between one-dimensional (1D) synthetic RIRs ($S$) and real RIRs ($R$) in the absence of paired training examples. Inspired by WaveGAN, which applies generative adversarial networks (GANs) to raw-waveform audio, we directly input RIRs as raw audio samples to our network to learn the mapping functions. In most cases, real and synthetic RIRs are less than one second in duration. Therefore, we re-sample the synthetic and real RIR datasets without loss of generality to 16 kHz and pass them as a one-dimensional input of length 16384.

We represent the real RIR training samples as $\{r_i\}_{i=1}^N$ where $r_i \in R$ and the synthetic RIR training samples as $\{s_i\}_{i=1}^N$, where $s_i \in S$. The data distributions of the training samples are $r \sim p_{data} (r)$ and $s \sim p_{data} (s)$. We use 2 generators to learn the mappings $G_{SR} : S \rightarrow R$ and $G_{RS} : R \rightarrow S$. Our goal is to learn the mapping $G_{SR} : S \rightarrow R$. We use the inverse mapping $G_{RS} : R \rightarrow S$ with cycle-consistency loss  \cite{cycle-loss} to preserve the acoustic characteristics in Synthetic RIRs during translation. We use discriminator $D_{R}$ to differentiate real RIRs  $\{r_i\}_{i=1}^N$ and translated synthetic RIRs $\{G_{SR}(s_i)\}_{i=1}^N$. Similarly, we use $D_{S}$ to discriminate $\{s_i\}_{i=1}^N$ and $\{G_{RS}(r_i)\}_{i=1}^N$. Our objective function consists of adversarial loss \cite{adversarial-loss}, cycle-consistency loss and identity loss \cite{identity-loss} to learn the mapping functions.

\subsubsection{Adversarial Loss}
To ensure the synthetic RIRs are translated to real RIRs, we use the following objective for the mapping function $G_{SR} : S \rightarrow R$ and the discriminator $D_{R}$.
{

\begin{equation}\label{ad_loss}
\begin{aligned}[b]
    \mathcal{L}_{adv}(G_{SR},D_{R},S,R) = \mathbb{E}_{r \sim p_{data}{(r)}}[\log{D_{R}(r)}] \\
    + \mathbb{E}_{s \sim p_{data}{(s)}}[\log(1 - {D_{R}(G_{SR}(s))}].
\end{aligned}
\end{equation}
}

\noindent The discriminator $D_{R}$ tries to distinguish between translated RIRs using the mapping function $G_{SR} : S \rightarrow R$ from the real RIRs by maximizing ($max$) the adversarial loss. The generator $G_{SR} : S \rightarrow R$ attempts to generate real RIRs that tend to minimize ($min$) the adversarial loss, i.e., $\min_{G_{SR}} \max_{D_{R}} \mathcal{L}_{adv}(G_{SR},D_{R},S,R)$. Similarly, we train the mapping function $G_{RS} : R \rightarrow S$ and the discriminator $D_{S}$ with the objective $\mathcal{L}_{adv}(G_{RS},D_{S},R,S)$.

\subsubsection{Cycle Consistency Loss}
We use cycle consistency loss to preserve the acoustic characteristics in the RIRs during the translation. The cycle consistency loss ensures that $G_{RS}(G_{SR}(s)) \sim s$ and $G_{SR}(G_{RS}(r)) \sim r$.

{
\begin{equation}\label{cy_loss}
\begin{aligned}[b]
    \mathcal{L}_{cyc}(G_{SR},G_{RS}) = \mathbb{E}_{s \sim p_{data}{(s)}}[||G_{RS}(G_{SR}(s)) - s ||_1] \\
    + \mathbb{E}_{r \sim p_{data}{(r)}}[||G_{SR}(G_{RS}(r)) - r ||_1].
\end{aligned}
\end{equation}
}

\subsubsection{Identity Mapping Loss}
Identity mapping loss preserves the amplitude of input RIRs:
{
\begin{equation}\label{id_loss}
\begin{aligned}[b]
    \mathcal{L}_{id}(G_{SR},G_{RS}) = \mathbb{E}_{s \sim p_{data}{(s)}}[||G_{RS}(s) - s ||_1] \\
    + \mathbb{E}_{r \sim p_{data}{(r)}}[||G_{SR}(r) - r ||_1].
\end{aligned}
\end{equation}
}

\subsubsection{Full Objective}
The overall objective function can be given as
{
\begin{equation}\label{full_loss}
\begin{aligned}[b]
    \mathcal{L}_(G_{SR},G_{RS},D_{S},D_{R}) = \mathcal{L}_{adv}(G_{SR},D_{R},S,R) \\
    + \mathcal{L}_{adv}(G_{RS},D_{S},R,S) \\
    + \lambda_{cyc} \mathcal{L}_{cyc}(G_{SR},G_{RS}) \\
    + \lambda_{id} \mathcal{L}_{id}(G_{SR},G_{RS}),
\end{aligned}
\end{equation}
}

\noindent where $\lambda_{cyc}$ and $\lambda_{id}$ control the relative importance of cycle consistency loss and identity mapping loss, respectively. We train our TS-RIRGAN to find the optimal mapping functions $G_{SR}^{*}$ and $G_{RS}^{*}$ by solving
{
\[
\begin{aligned}[b]
    G_{SR}^{*}, G_{RS}^{*} = \arg \min_{G_{SR},G_{RS}} \max_{D_{S},D_{R}} \mathcal{L}(G_{SR},G_{RS},D_{S},D_{R}).
\end{aligned}
\]
}
We use $G_{SR}^{*}$ to translate imprecise synthetic RIRs to real RIRs.

\subsubsection{Implementation}

\textbf{Network Architecture: } We adapt the discriminator architecture from Donahue et al.~\cite{WaveGAN}. We did not use the phase shuffle operation proposed in WaveGAN~\cite{WaveGAN} because this operation did not improve our results. Inspired by Johnson et al.~\cite{Generator}, we designed our generator network consisting of an encoder, a transformer and a decoder. Figure \ref{architecture} highlights our generator and discriminator architectures. Similar to WaveGAN, we use 1D filters of length 25 to perform convolution and transposed convolution operations in our TS-RIRGAN architecture.

\textbf{Dataset:} We use an equal number of real RIRs from BUT ReverbDB \cite{ButReverb} and synthetic RIRs generated using the geometric acoustic simulator \cite{zhenyu_GAS} to train our TS-RIRGAN architecture. The BUT ReverbDB consists of 1891 RIRs covering the office, hotel room, conference room, lecture room, meeting room, and stairs. We remove repeated RIRs and RIRs recorded in stairs. Among 1209 retained RIRs in BUT ReverbDB, we train our network using 967 RIRs and keep 242 RIRs for testing purposes. Room dimensions, loudspeaker location, and microphone location corresponding to each real RIRs are documented in BUT ReverbDB dataset. We use this information to generate synthetic RIRs using the geometric acoustic simulator. We use random surface absorption/reflection coefficients to generate synthetic RIRs because we do not have room-material information. Therefore one-to-one mapping between synthetic and real RIRs should not be expected.


\subsection{Sub-band Room Equalization (EQ)}
Sub-band room equalization bridges the gap in the frequency gain of real and synthetic RIRs over the entire frequency range. Our formulation is based on the sub-band room equalization approach described in \cite{zhenyu_low}. Sub-band relative gain calculation and equalization matching are the two stages in sub-band room equalization.

\subsubsection{Sub-band relative gain calculation}
We calculate the re-sampled relative gains to compensate for the difference in relative gains between synthetic and real RIRs. We compute the frequency response of every RIR in a real-world dataset \cite{ButReverb}. We compute the relative gain from the frequency response by taking the gain at 1000Hz as the reference for each real RIR. Then we extract the relative frequency gain at 7 unique sample points (62.5Hz, 125Hz, 250Hz, 500Hz, 2000Hz, 4000Hz, 8000Hz) for every real RIR. The mean and standard deviations of the relative gains for each sample point are different. Therefore we use a Gaussian mixture model to model 7 Gaussian distributions using the relative gains from the sampled points. We re-sample equal numbers of relative gains for each sample point as the input to the Gaussian mixture model. Instead of using the relative gains of the real RIRs, we use the re-sampled relative gains. We use re-sampled relative gains to avoid duplicating the real RIRs during equalization matching. We choose the reference and the number of sample points as proposed in \cite{zhenyu_low}.

\subsubsection{Equalization matching}
We match the relative gains of synthetic RIRs with the re-sampled relative gains calculated from real RIRs. We compute the relative frequency gains for the synthetic RIRs at the chosen sample points (62.5Hz, 125Hz, 250Hz, 500Hz, 2000Hz, 4000Hz, 8000Hz), taking gain at 1000Hz as the reference. We calculate the difference in the relative gains of synthetic RIRs and the re-sampled relative gains. Next, we design a finite impulse response (FIR) filter using the window method \cite{window} to compensate for the difference in the relative gains. We filter the synthetic RIRs using our designed FIR filter to match the sub-band relative gains of synthetic RIRs with the re-sampled relative gains.

\begin{table}[t]
    \setlength{\tabcolsep}{3pt}
	\caption{Different combinations of our post-processing methods studied in this paper. The best combination is marked in \textbf{bold}.}
	\label{table:combination}
	\centering
	\begin{tabular}{@{}llr@{}}	
		\toprule
		\textbf{Combination} & \textbf{Description} \\ 
		\midrule
		GAS+EQ & Only perform room equalization.\\ 
		\midrule
		$G_{SR}^{*}$(GAS+EQ) & First, perform room equalization,   \\ 
		& then translate  the equalized synthetic RIR \\ 
		& to a real RIR.\\
		\midrule
		$G_{SR}^{*}$(GAS) & Only translate synthetic RIR to real RIR.\\ 
		\midrule
		 \textbf{\boldmath $G_{SR}^{*}$(GAS)+EQ} &  \textbf{First, translate a synthetic RIR to a real}  \\
		& \textbf{RIR, then perform room equalization}\\
		& \textbf{ to the translated RIR.}\\
		\bottomrule
	\end{tabular}
\end{table}

\begin{table*}[t]
   \setlength{\tabcolsep}{8pt}
	\caption{Mean values of the acoustic parameters. We calculated the mean reverberation time ($T_{60}$), mean direct-to-reverberant ratio (DRR), mean early-decay-time (EDT), and mean early-to-late index (CTE) for real, synthetic and post-processed synthetic RIRs. We also report the absolute mean difference of the acoustic parameters between synthetic and post-processed synthetic RIRs and real RIRs. The acoustic parameter values with the least absolute mean difference are shown in \textbf{bold}. }
	\label{table:statistics}
	\centering
	\begin{tabular}{@{}llllllllr@{}}	
		\toprule
		\textbf{RIRs} & \multicolumn{2}{c}{{\textbf{\boldmath$T_{60}$ (seconds)} }}
			& \multicolumn{2}{c}{{\textbf{DRR (dB)} }}&\multicolumn{2}{c}{{\textbf{EDT (seconds)}}}	& \multicolumn{2}{c}{{\textbf{CTE (dB)}}}\\
\cmidrule(r{4pt}){2-9} 
		& \textbf{Mean} & \textbf{Difference}& \textbf{Mean} & \textbf{Difference}& \textbf{Mean} & \textbf{Difference}& \textbf{Mean} & \textbf{Difference} \\  
		\midrule
		Real RIRs& 1.0207 &  &-6.3945& &0.8572&&3.4886&\\
		GAS& \textbf{0.9553}&\textbf{0.0654}&-4.7277&1.6668&0.8846&0.0274&4.7536&1.265\\  
		GAS+EQ& 0.9540&	0.0667&-7.4246&1.0301&0.8912&	0.0340&	5.6404&	2.1518\\
		$G_{SR}^{*}$(GAS+EQ)& 1.5493&0.5286&-8.3879&1.9934&1.046&0.1888&2.6562&	0.8324\\   

		$G_{SR}^{*}$(GAS)& 1.6433&0.6226&\textbf{-6.6491}&\textbf{0.2546}&\textbf{0.8483}&\textbf{0.0089}&\textbf{3.4907}&\textbf{0.0021}\\
		$G_{SR}^{*}$(GAS)+EQ& 1.6364&0.6157&-6.7234&0.3289&0.8323&0.0249&3.5367&0.0481\\
			
		\bottomrule
	\end{tabular}
\end{table*} 

\begin{figure}[t] 
\centering
\subfloat[GAS.]{\includegraphics[width=0.46\columnwidth]{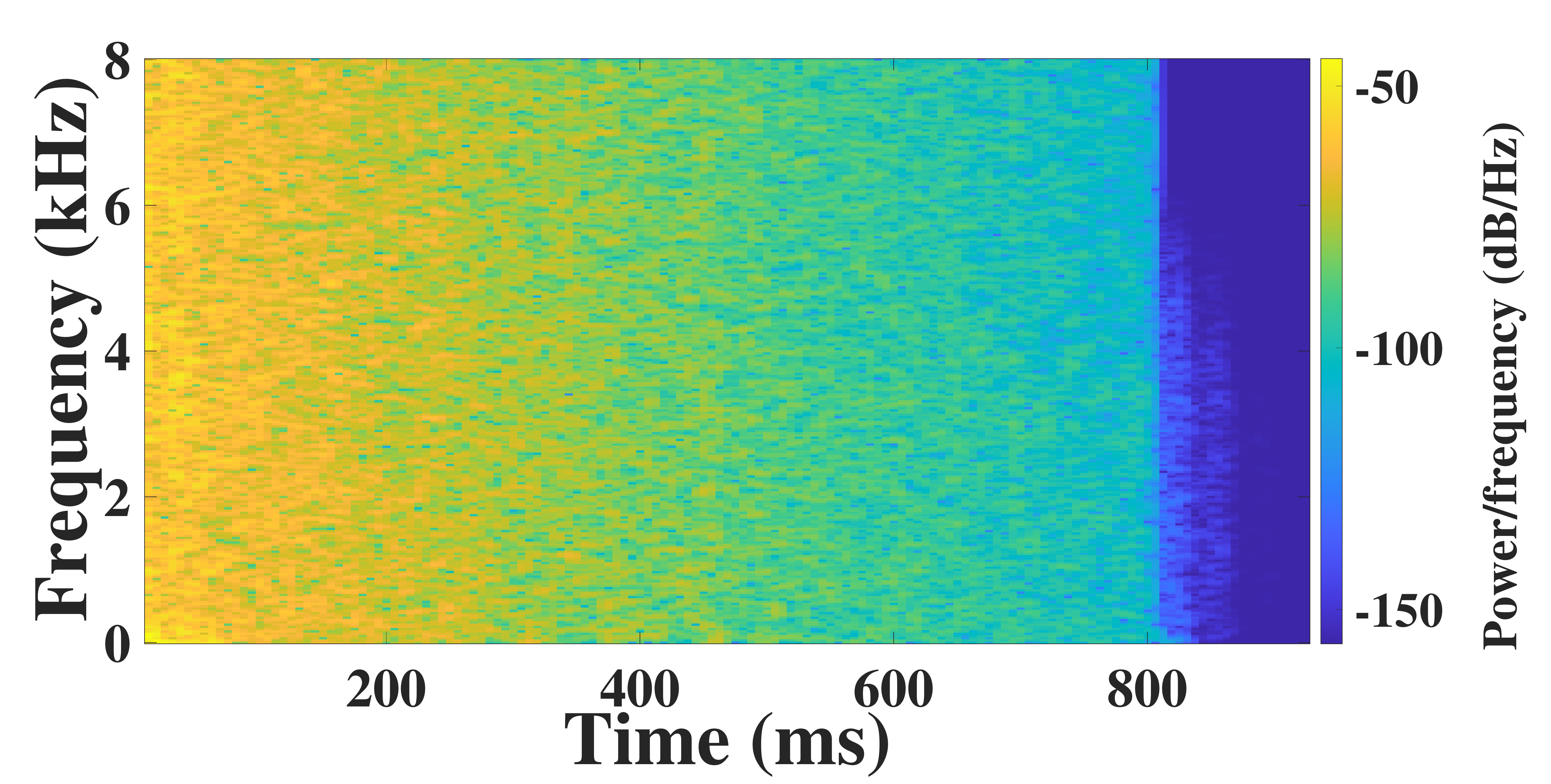} \label{GAS}}
\quad
\subfloat[GAS+EQ.]{\includegraphics[width=0.46\columnwidth]{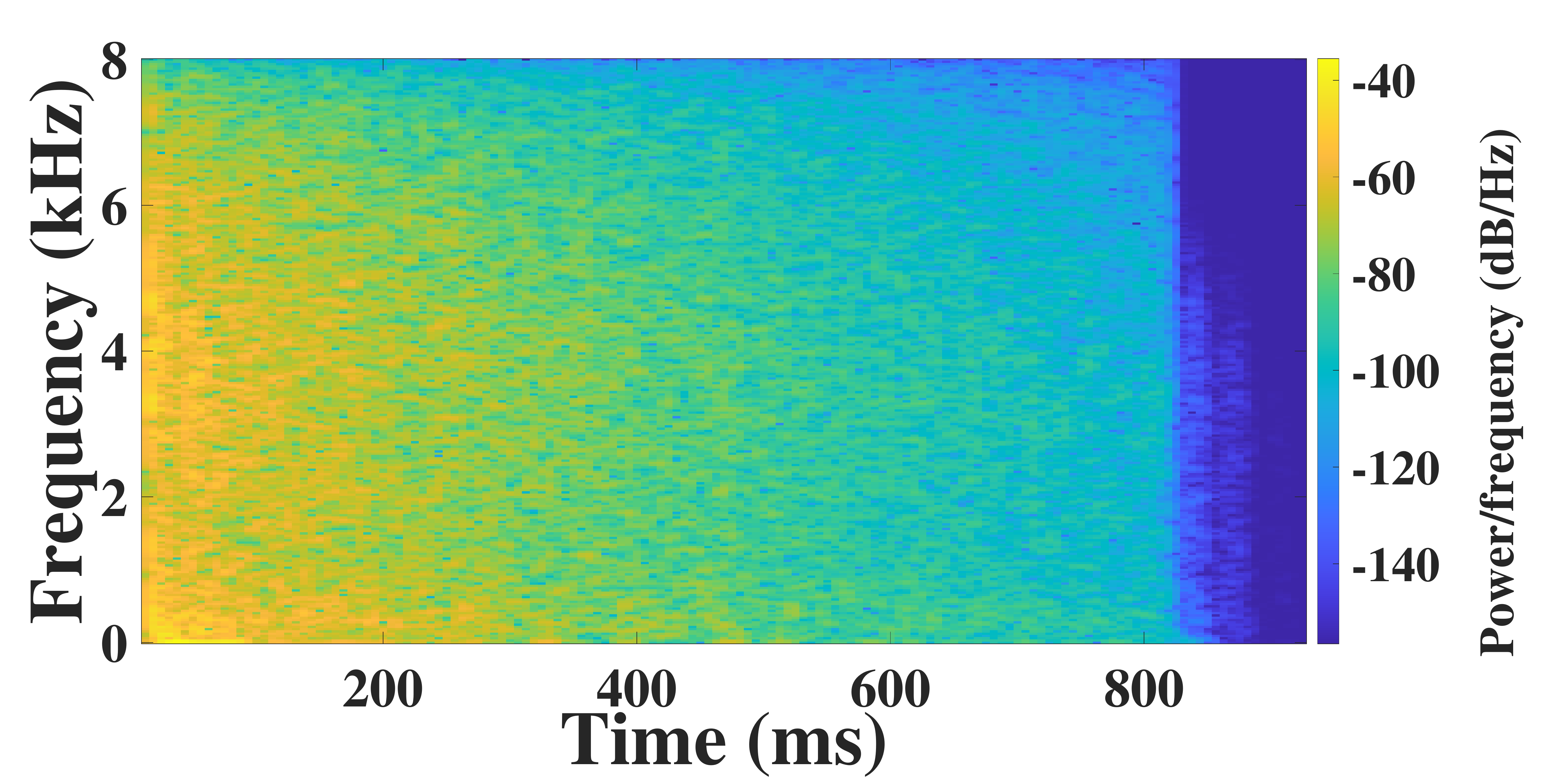}\label{GAS+EQ}}
\quad
\subfloat[$G_{SR}^{*}$(GAS+EQ).]{\includegraphics[width=0.46\columnwidth]{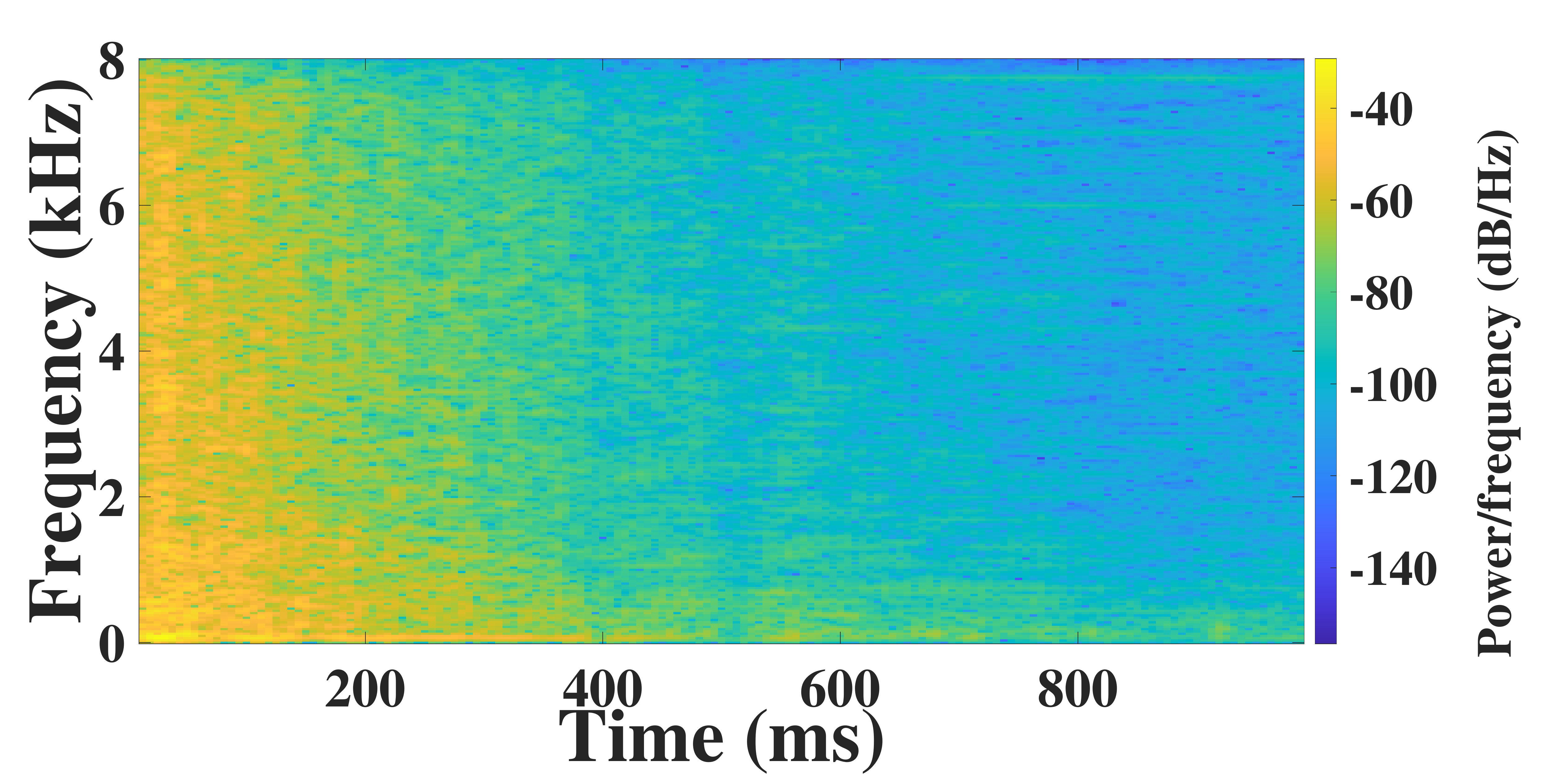}\label{T(GAS+EQ)}}
\quad
\subfloat[$G_{SR}^{*}$(GAS).]{\includegraphics[width=0.46\columnwidth]{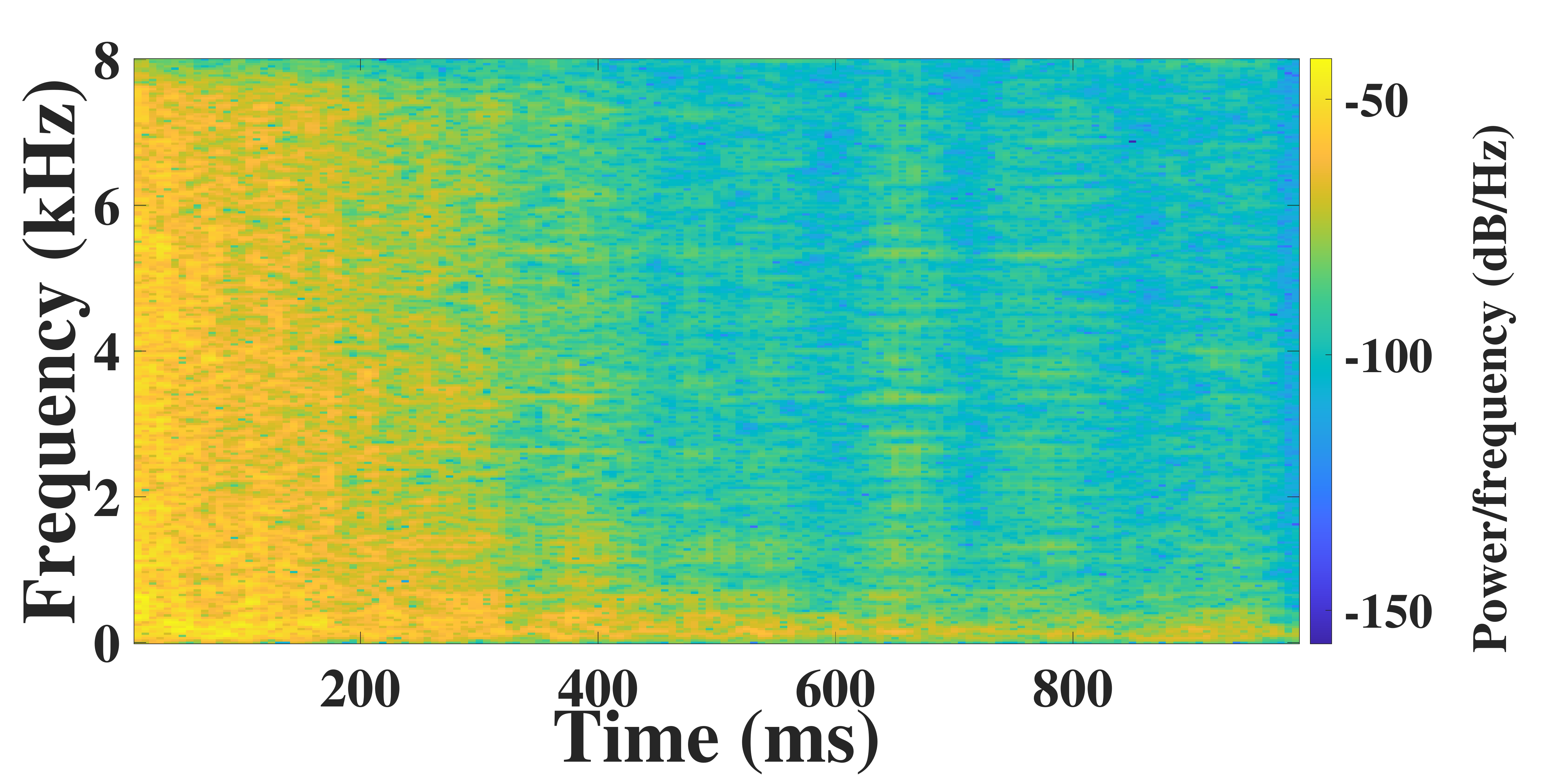}\label{T(GAS)}}
\quad
\subfloat[$G_{SR}^{*}$(GAS)+EQ.]{\includegraphics[width=0.46\columnwidth]{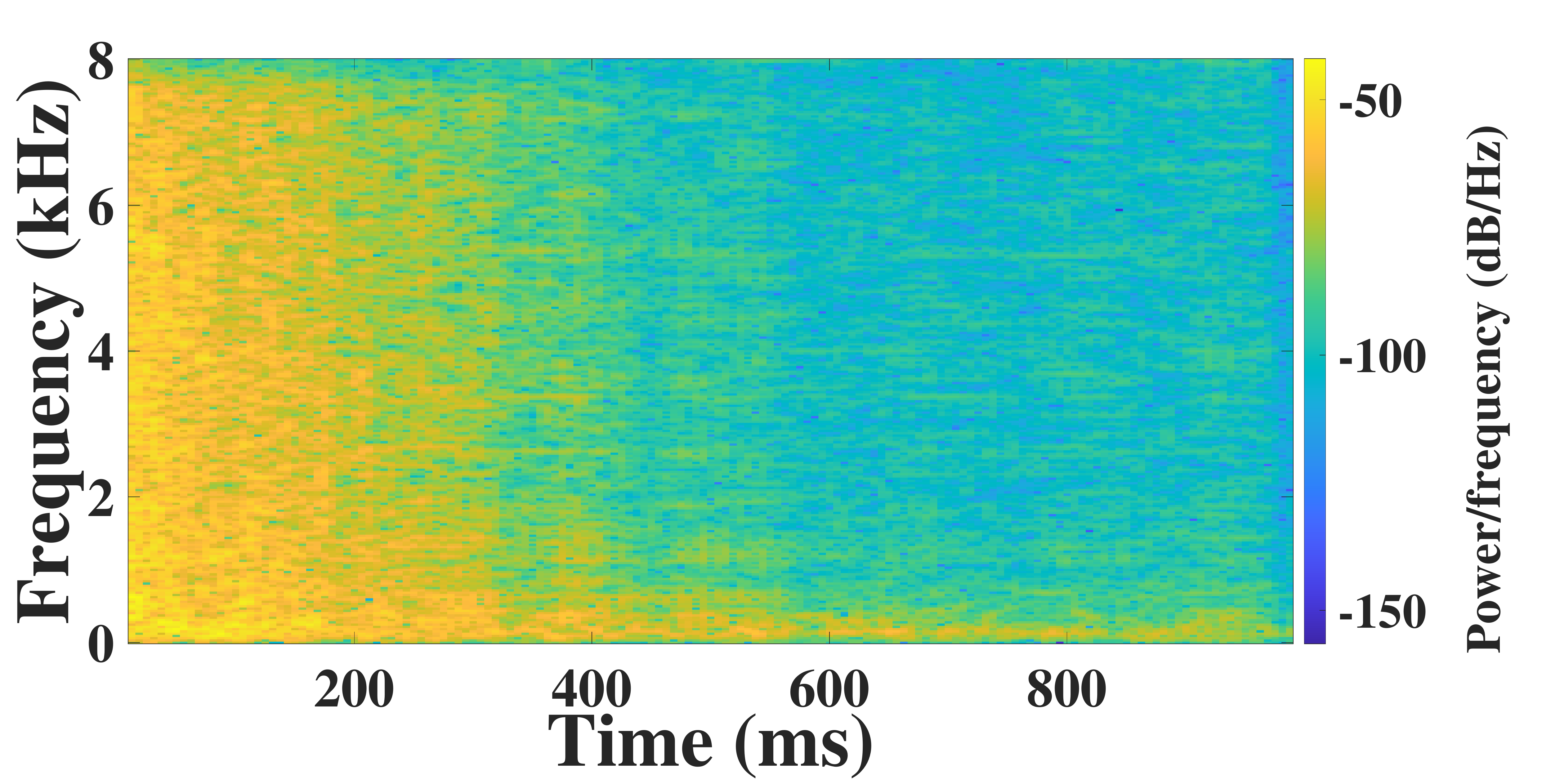}\label{T(GAS)+EQ}}
\quad
\subfloat[Real RIR.]{\includegraphics[width=0.46\columnwidth]{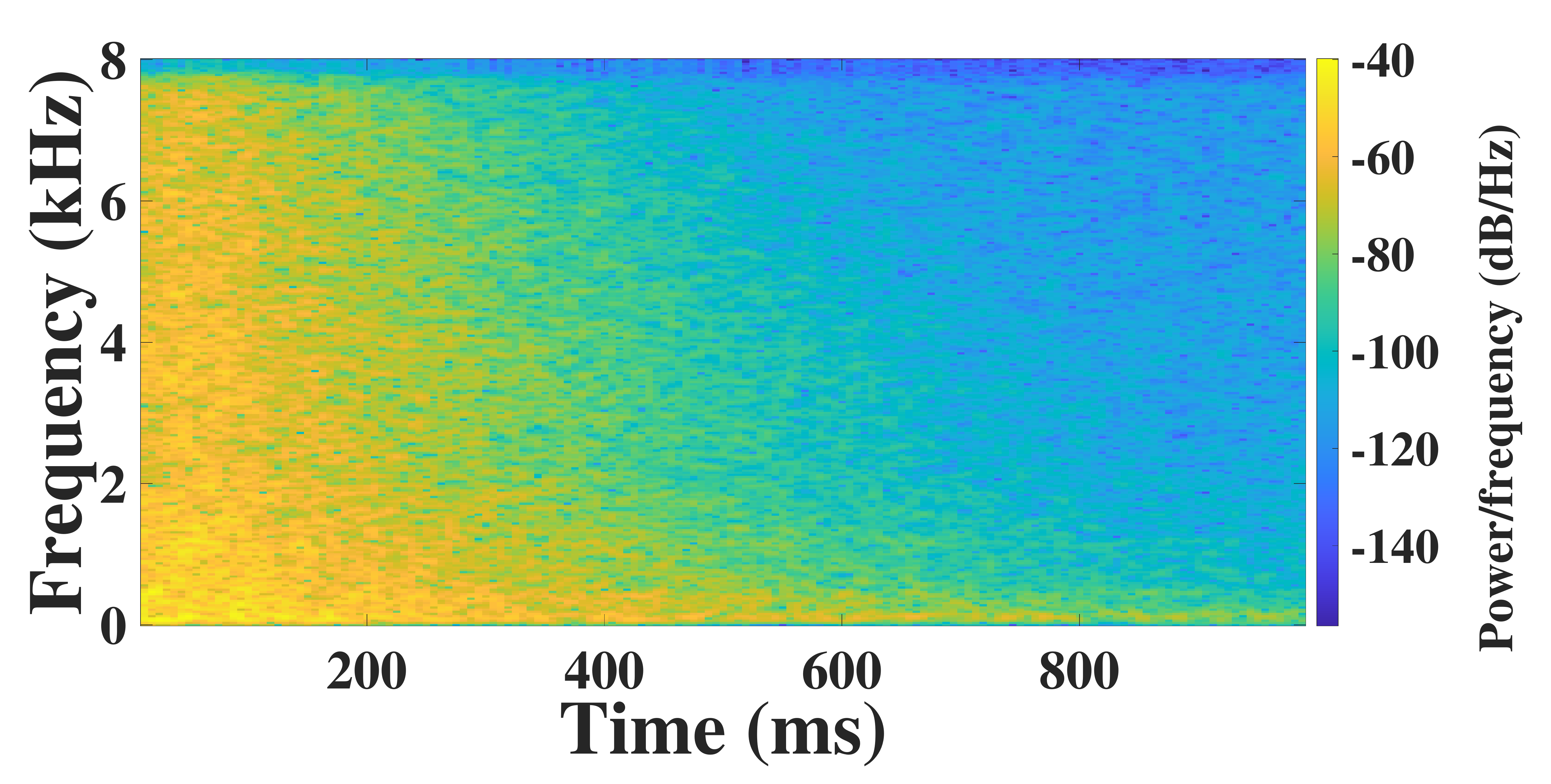}\label{Real}}
\caption{The spectrogram of a synthetic RIR generated using the geometric acoustic simulator \cite{zhenyu_GAS} (Figure \ref{GAS}), post-processed synthetic RIRs (Figure \ref{GAS+EQ}-\ref{T(GAS)+EQ}), and a real RIR (Figure \ref{Real}). Sub-band room equalization (EQ) and synthetic RIR to real RIR translation ($G_{SR}^{*}$()) are the two methods used to post-process the synthetic RIR in different combinations (Table \ref{table:combination}). Among the energy distribution in spectrograms of post-processed synthetic RIRs, the energy distribution in $G_{SR}^{*}$(GAS)+EQ is closest to the spectrogram of a real RIR. We can observe that the energy distribution over the low-frequency and high-frequency region in $G_{SR}^{*}$(GAS)+EQ is similar to a real RIR.}
\label{spectrogram}
\end{figure}

\subsection{Optimal Combination}
We tried different combinations (Table \ref{table:combination}) of our post-processing approach to come up with the optimal combination. We estimated 4 different acoustic parameter values from synthetic RIRs generated using the geometric acoustic simulator \cite{zhenyu_GAS} (GAS), post-processed synthetic RIRs using a different combination of our post-processing approach and real RIRs to evaluate how much the post-processed RIRs are closer to real RIRs. Reverberation time ($T_{60}$), direct-to-reverberant ratio (DRR), early-decay-time (EDT), and early-to-late index (CTE) are four acoustic parameters used for our evaluation. $T_{60}$ is the time required to decay the sound pressure by 60 decibels (dB). The ratio of the sound pressure level of a direct sound source to the sound pressure level of reflected sound in dB is called DRR \cite{drr_book}. EDT is calculated by multiplying the time taken for the sound source to decay by 10 dB by a factor of 6.  The proportion of the total sound energy received in the first 50ms to the energy received during the rest of the period is called CTE \cite{room_acoustics_vigran_2014}. Synthetic RIRs and real RIRs used to train TS-RIRGAN and to perform sub-band room equalization do not have one-to-one mapping. Therefore, we calculated the mean values for different acoustic parameters to evaluate our post-processing approach.

Table \ref{table:statistics} presents the mean values of the acoustic parameters for different sets of RIRs and the absolute difference between the mean values of the acoustic parameters of synthetic and post-processed synthetic RIRs and real RIRs. We can see that mean DRR, mean EDT, and mean CTE values of $G_{SR}^{*}$(GAS) and $G_{SR}^{*}$(GAS)+EQ are closer to the real RIRs when compared with the other combinations of our post-processing approach. Therefore our proposed TS-RIRGAN is capable of improving the quality of synthetic RIRs by translating the wave effects present in real RIRs to synthetic RIRs. However, we can see a deviation in the mean $T_{60}$ values for the post-processed synthetic RIRs using TS-RIRGAN.

\begin{figure}[t] 
	\centering
    \includegraphics[width=3.3in]{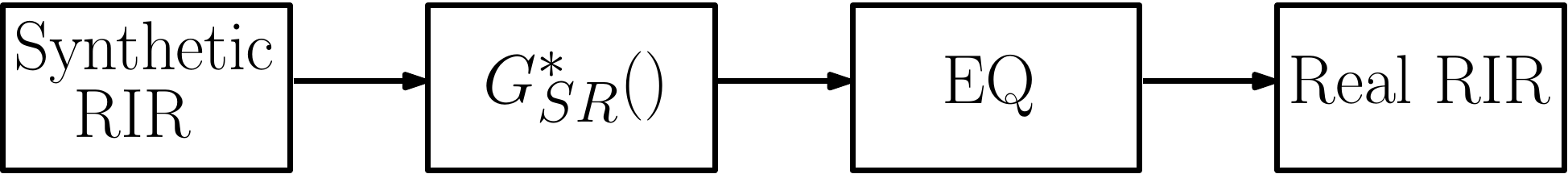}
	\caption{Our overall pipeline to improve the quality of synthetic RIRs. We translate synthetic RIRs to real RIRs using our learned mapping function $G_{SR}^{*}()$, then we augment the wave effects in translated synthetic RIRs by performing real-world sub-band room equalization (EQ).}
	\label{optimal}
\end{figure}

Figure \ref{spectrogram} shows the spectrogram of a synthetic RIR generated using the geometric acoustic simulator \cite{zhenyu_GAS} (GAS), post-processed synthetic RIRs using a different combination of our post-processing approach and a real RIR. From the spectrograms, we can see that by translating a synthetic RIR to a real RIR, we improve the energy distribution in the low-frequency region (Figure \ref{T(GAS)}) by compensating low-frequency wave effects present in real RIRs. When we perform sub-band room equalization after translation, we observe further refinement in the spectrogram (Figure \ref{T(GAS)+EQ}), especially around 600ms to 800ms. After trying all the combinations, we highlight the optimal combination in Figure \ref{optimal}. We chose the optimal combination based on the set of acoustic parameter values and the energy distribution of the post-processed RIRs.

\section{Implementation and Results}

\subsection{Benchmark}
We evaluate our approach on the Kaldi LibriSpeech far-field ASR recipe \cite{zhenyu_low}. We convolve clean speech $x_c[t]$ from LibriSpeech \cite{LibriSpeech} with different sets of RIRs $r[t]$ and add environmental noise $n[t]$ from BUT ReverbDB \cite{ButReverb} to augment a far-field speech $x_f[t]$ training dataset. The environmental noise is started at a random position $l$ and repeated in a loop to fill the clean speech. In Equation \ref{eq:conv}, $\lambda$ is calculated for different signal-to-noise ratios, which ranges from 1dB to 2dB:
{
\begin{equation}\label{eq:conv}
\begin{aligned}[b]
x_f[t] =  x_c[t] \circledast r[t] + \lambda * n[t+l].  
\end{aligned}
\end{equation}
}

We train time-delay neural networks \cite{Kaldi_network} using our augmented training dataset. After training the network, we decode the identity vectors \cite{ivector} (i-vectors) of a far-field speech test set using phone language models. We calculate word error rate (WER) for large four-gram (fglarge), large tri-gram (tglarge), medium tri-gram (tgmed), and small tri-gram (tgsmall) phone language models, as well as online decoding using a tgsmall phone language models. During online decoding, the i-vectors extracted from the far-field speech test set are passed in real-time. We use WER to evaluate the far-field speech augmented using different sets of RIRs.

Training and testing on the benchmark for each far-field speech training dataset take around 4 days. We ran all the experiments in the same environment to perform a fair comparison.
\subsection{Data Preparation}
We use real RIRs and environmental noise from BUT ReverbDB \cite{ButReverb} and clean speech (test-clean) from LibriSpeech \cite{LibriSpeech} to augment a real-world far-field speech test set using Equation \ref{eq:conv}. We evaluate our proposed method using the real-world far-field speech test set. We randomly split 1209 RIRs in BUT ReverbDB \cite{ButReverb} into subsets of \{773,194,242\} to create training, development, and test far-field speech datasets. 

We use the meta-info accompanying each real RIR to generate synthetic RIRs using the state-of-the-art geometric acoustic simulator (GAS). We post-process the synthetic RIRs by translating synthetic RIRs to real RIRs and performing real-world sub-band room equalization in different combinations (Table \ref{table:combination}). 

We also generated RIRs using the pre-trained IR-GAN \cite{irgan} on BUT ReverbDB dataset \footnote{\url{https://gamma.umd.edu/pro/speech/ir-gan}}. IR-GAN is a neural network based RIR generator that can generate realistic RIRs corresponding to different acoustic environment by parametrically controlling acoustic parameters. 

We created different far-field speech training set by convolving LibriSpeech training datasets (train-clean-\{100,360\}) with different RIRs and adding environmental noise from BUT ReverbDB \cite{ButReverb} set using Equation \ref{eq:conv}. We use synthetic RIRs generated using GAS, post-processed synthetic RIRs, RIRs generated using IR-GAN and real RIRs to augment different training far-field speech datasets.



\begin{table}[t]
   \setlength{\tabcolsep}{1.8pt}
	\renewcommand{\arraystretch}{0.85} 
	\caption{Word error rate (WER) reported by the Kaldi far-field ASR system. We trained the Kaldi model using the different augmented far-field speech training sets and tested it on a real-world far-field speech. The training sets are augmented using synthetic RIRs generated using GAS, post-processed synthetic RIRs (Table \ref{table:combination}), synthetic RIRs generated using IR-GAN and real RIRs. We report WER for fglarge, tglarge, tgmed, and tgsmall phone language models and online decoding using tgsmall phone language model. Our best results are shown in \textbf{bold}.}
	\label{table:results}
	\centering
	\begin{tabular}{@{}llllllr@{}}	
		\toprule
		\multicolumn{2}{l}{\multirow{2}{24mm}{\textbf{Training data}}}
			& \multicolumn{5}{c}{\textbf{Test Word Error Rate (WER)} [\%]}\\
\cmidrule(r{4pt}){3-7} 
		&	& fglarge & tglarge & tgmed & tgsmall & online\\  
		\midrule
		&clean (Baseline) & 77.15 & 77.37 & 78.00 & 78.94 & 79.00\\
		&real (Oracle) & 12.40 & 13.19 & 15.62 & 16.92 & 16.88\\  
		\midrule
		&GAS\cite{zhenyu_GAS}  & 16.53 & 17.26 & 20.24 & 21.91 & 21.83\\
		&GAS+EQ\cite{zhenyu_low}  & 14.51 & 15.37 & 18.33 & 20.01 & 19.99\\   
		\midrule
		&$G_{SR}^{*}$(GAS+EQ)   & 14.27 & 14.98 & 17.79 & 19.37 & 19.36\\
		\textbf{Ours}& $G_{SR}^{*}$(GAS)   & 14.12 & 14.70 & 17.44 & 19.08 & 19.06\\
		&\textbf{\boldmath $G_{SR}^{*}$(GAS)+EQ} & \textbf{13.24} & \textbf{14.04} & \textbf{16.65} & \textbf{18.40} & \textbf{18.39}\\
		\midrule
		&GAS\cite{zhenyu_GAS}  & 16.53 & 17.26 & 20.24 & 21.91 & 21.83\\
		&IR-GAN\cite{irgan}  & 14.99 & 15.93 & 18.81 & 20.28 & 20.24\\
		&GAS+IR-GAN\cite{irgan}  & 14.16 & 14.99 & 17.56 & 19.21 & 19.21\\
		\textbf{Ours} &\textbf{\boldmath $G_{SR}^{*}$(GAS)+EQ}  & \textbf{13.24} & \textbf{14.04} & \textbf{16.65} & \textbf{18.40} & \textbf{18.39}\\		
		\bottomrule
	\end{tabular}
\end{table} 

\subsection{Results and Analysis}
Table \ref{table:results} shows the word error rate (WER) reported by the Kaldi LibriSpeech far-field ASR benchmark \cite{zhenyu_low}. We can see that the augmented far-field speech training sets perform well compared to our baseline model trained on a clean Librispeech dataset. The lowest WER is reported by our oracle model trained on real-world far-field speech. In our work, we aim to minimize the gap in the performance between real RIRs and synthetic RIRs.


The WERs for tgsmall reported by GAS+EQ and $G_{SR}^{*}$(GAS) are 18.33\% and 17.44\%, respectively. We observe that our approach outperforms the prior methods by up to 4.8\%. We see an interesting observation with $G_{SR}^{*}$(GAS+EQ) and $G_{SR}^{*}$(GAS) datasets. When compared to translated synthetic RIRs ($G_{SR}^{*}$(GAS)), translated room equalized RIRs ($G_{SR}^{*}$(GAS+EQ)) perform poorly. 

\textbf{Optimal Approach: } We can see that translating imprecise synthetic RIRs to real RIRs and performing real-world sub-band room equalization on the translated RIRs ($G_{SR}^{*}$(GAS)+EQ) gives the lowest WER. When compared to training sets created using unmodified RIRs (GAS) and room equalized RIRs (GAS+EQ), we observe a relative reduction in WER by up to 19.9\% and 9.1\%, respectively. 

Physical-based acoustic simulators (GAS) and neural-network-based RIR generators (IR-GAN) generate RIRs using two different approaches. GAS models RIR corresponding to a particular scene by considering room dimension, speaker, listener position, etc. IR-GAN uses acoustic parameters to generate an RIR for a particular scene. In previous work, \cite{irgan}, far-field speech augmented using synthetic RIRs from GAS and IR-GAN are used to train a robust far-field ASR system (GAS+IR-GAN). From Table \ref{table:results}, we can observe that our post-processed RIRs using our optimal approach ($G_{SR}^{*}$(GAS)+EQ) outperforms the combination of RIRs generated using GAS and IR-GAN (GAS+IR-GAN).


\section{Conclusion}
We present a new architecture to translate synthetic RIRs to real RIRs and perform real-world sub-band room equalization on the translated RIRs to improve the quality of synthetic RIRs. We evaluate the quality of our post-processed synthetic RIRs using a set of acoustic parameter values and the energy distribution of the post-processed RIRs. The set of  acoustic parameter values indicates how much the wave effects in post-processed RIRs are closer to real RIRs. We show that the mean direct-to-reverberant ratio, mean early-decay-time, and mean early-to-late index of the post-processed synthetic RIRs are closer to the real RIRs when compared to the unmodified synthetic RIRs. We also evaluate our post-processing approach on the Kaldi LibriSpeech far-field automatic speech recognition benchmark and observe that our post-processed RIRs outperform unmodified synthetic RIRs by up to 19.9\%. In the future, we would like to explore improving the quality of synthetic RIRs based on improved techniques to model acoustic wave effects and translation architectures. We would also like to evaluate their benefits for other applications, including speech separation~\cite{cone-of-silence,aralikatti} and audio-visual speech recognition~\cite{Audio-Visual} tasks.

\section{Acknowledgements}
This work is supported in part by ARO grant W911NF-18-1-0313,  NSF grant \#1910940,  Capital One and Intel.

\bibliographystyle{IEEEbib}
\bibliography{strings,refs}

\end{document}